\documentclass[12pt]{iopart}

\usepackage{graphicx,amsfonts,amsbsy, amssymb, amsthm,  mathrsfs}

\eqnobysec
\newcommand{\eqref}[1]{(\ref{#1})}

\begin{document}
\title{Exterior - Interior Duality for Discrete Graphs}
\author{Uzy Smilansky}
\address{Department of Physics of Complex Systems,
Weizmann Institute of Science, Rehovot 76100, Israel.}
\address{School of Mathematics, Cardiff University, Cardiff,
Wales,UK}
 \ead{\mailto{uzy.smilansky@weizmann.ac.il}}
\begin{abstract}
The Exterior-Interior duality expresses a deep connection between
the Laplace spectrum in bounded and  connected  domains in
$\mathbb{R}^2$, and the scattering matrices in the exterior of the
domains. Here, this link is extended to the study of the spectrum of
the discrete Laplacian on finite graphs. For this purpose, two
methods are devised for associating scattering matrices to the
graphs. The Exterior -Interior duality is derived for both methods.
\end{abstract}
\section{Introduction}\label{sec:intro}
The purpose of the present paper is to introduce to the study of the
Laplacian on discrete graphs a concept which found several
applications in the spectral theory of the wave (and the
Schr\"odinger) operator  - the exterior-interior duality. It applies
for wave equations on manifolds which can be partitioned into
interior (assumed to be simply connected and compact) and exterior
domains separated by a boundary. One considers now the wave
equations in the two domains,  subject to the same boundary
conditions. The wave operator restricted to the interior has a pure
point spectrum, while in the exterior the spectrum is continuous and
a scattering operator can be defined. The restriction of the
scattering operator to a particular value of the spectral parameter
is the unitary (on shell) scattering matrix $S(\lambda)$. Its
spectrum is confined to the unit circle. The exterior-interior
duality asserts that the spectrum of the interior can be identified
as those values of $\lambda$ for which one of the eigenvalues of the
$S$ matrix approaches the value $1$.

The exterior-interior duality was first discussed in the physics
literature in \cite{doron1,doron2,dietz,leshouches}. It was
presented in several versions, which differ in  the geometry of the
space where the scattering operator is defined.  Consider e.g., a
compact domain $\Omega  \subset \mathbb{R}^2$ and the wave equation
subject to Dirichlet boundary conditions. In the exterior, the wave
equation is subject to the same boundary conditions, and $\Omega$ is
treated as an obstacle. The scattering it induces is described in
terms of the scattering matrix $S(\lambda)$. The exterior-interior
duality ensures that the spectrum of  the Dirichlet Laplacian in the
interior is identified by the values of the spectral parameters
where an eigenvalue of $S(\lambda)$ approaches unity. This was
precisely formulated and proved in \cite{EP1,EP2,DEPSU}.

\begin{figure*}%
  \begin{center}%
    \includegraphics[width={.6 \linewidth}]
    {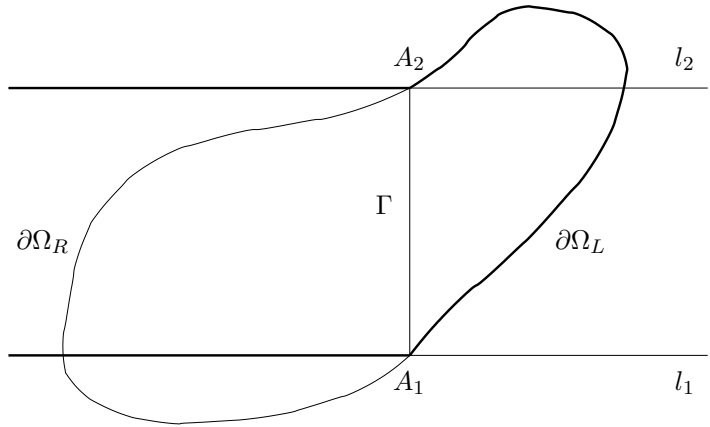}
    \caption{%
The domain and the added auxiliary channels which are used for the
definition of the scattering matrix.}
\label{fig:int-ext-billiard}%
\end{center}%
\end{figure*}%

The wave-guide version which is closer to the present discussion can
be crudely described as follows: Consider a connected domain in
$\mathbb{R}^2$ and impose Dirichlet boundary conditions on its
boundary $\partial\Omega$ which is assumed to be smooth. The
Dirichlet Laplacian has a discrete spectrum. To define an associated
scattering system, choose on the boundary two arbitrary points $A_1$
and $A_2$ (see figure (\ref {fig:int-ext-billiard})). Draw two
parallel lines, $l_1$ through $A_1$ and $l_2$ through $A_2$ so that
$l_{1}$ and $l_{2}$ are perpendicular to the  chord
$\Gamma=(A_2,A_1)$.  Consider now the part of the boundary
$\partial\Omega_L$ which starts at $A_1$ and is traversed in the
positive mathematical sense as one goes from $A_1$ to $A_2$. The
domain which is bounded by $\partial \Omega_L$ and by the half lines
of $l_1 $ and  $l_2$ to the left of $\Gamma$ is a semi-infinite
channel,  terminated at one of its ends by $\partial\Omega_L$. This
domain will be denoted by $L$ (for ``left") and its boundaries are
drawn by darker lines. Impose Dirichlet conditions on the boundary
of $L$ and compute the scattering matrix $S_L(\lambda)$. Repeat the
same construction for $\partial\Omega_R$ which is the part of the
boundary which completes $\partial\Omega_L$ to the full boundary.
The scattering matrix $S_R(\lambda)$ in the ``right" channel $R$ is
defined in the same way as above (The domain boundaries are drawn
with lighter lines). The spectrum of the Laplacian in $\Omega$ is
obtained by solving the secular equation
\begin{equation}
\det \left(I-S(\lambda)\right )=0\ ; S(\lambda)= S_L
(\lambda)S_R(\lambda)
 \label{eq:originalEID}
\end{equation}
This equation expresses the essence of the exterior-interior
duality: the spectral information in the interior is obtained from
the scattering matrix computed for an appropriate exterior problem.
This version and its variations was studied in the physics
literature \cite{doron1,schanz,rouvinez}. Only a few examples were
studied with mathematical rigor \cite {bruning}.

The exterior-interior Duality was discussed and proved for quantum
graphs. \cite{KS1,KS2}.

Secular equation of the type (\ref{eq:originalEID}) have several
theoretical and practical advantages which were used for various
purposes in the study of the wave equation. In the sequel, analogous
secular equations for the spectrum of the discrete Laplacian on
graphs will be derived, thus extending the interior-exterior duality
to the discrete case. There are two conceptual obstacles which
should be cleared in order to carry out this program:

\noindent \emph{i.} Given a finite graph with a general
connectivity, there is no natural way to convert it to a scattering
problem analogous to the first version discussed above.

\noindent \emph{ii.} There is no natural definition for a
``boundary" on a compact graph, and if a boundary is defined, the
meaning of boundary conditions has to be explained.

In the following section the above difficulties will be addressed,
and the scattering matrices $S(\lambda)$ will be constructed in two
different settings. The exterior-interior duality will be realized
by deriving the corresponding secular equations of the form
(\ref{eq:originalEID}) for the spectrum of the graph.

A few applications will be presented in section
 (\ref{sect:applications}).

The following notations and concepts will be used throughout the
present work. A graph $\mathcal{G}$ is defined in terms of its
vertex set $\mathcal{V}$ and edge (bond) set $\mathcal{E}$. The
cardinality of these sets will be denoted by $V=|\mathcal{V}|\ ,\
E=|\mathcal{E}|$. The graph topology is defined by the connectivity
(adjacency) matrix $C$ which is labeled by the vertex indices. It is
a symmetric matrix whose $i,j$ element counts the number of edges
connecting the vertices $i$ and $j$. The valency (degree) of a
vertex is defined as the number of bonds which emanate from it,
$v_i=\sum_j C_{i,j}$. The diagonal matrix $D={\rm diag}(v_i)$
appears in the definition of the graph Laplacian:
\begin{equation}
\Delta = -C  +D \ .
 \label{eq:lap0}\
\end{equation}

As explained above, and in a complete analogy with the example
discussed in the introduction, the graph whose spectrum one wishes
to compute will be extended so that it forms the ``interior graph"
$\mathcal{G}^{(0)}$ in a scattering setting.  All the quantities
which are related to the ``interior" graph will be denoted by a
superscript $0$. For simplicity it is assumed to be connected and
devoid of parallel bonds ($C^{(0)}_{i,j} \in
 \{0,1\}$) or loops ($C^{(0)}_{i,i}=0$).

\section{Scattering on discrete graphs}
\label{sec:scat} In this section  scattering on graphs will be
defined using two different approaches. The first goes along the
lines of \cite{Boris}, the second relies on the graph evolution
operator defined in \cite{US07}. For the sake of completeness, some
background material for each of the methods will be reviewed before
the scattering matrices are presented.
\subsection{Scattering into attached leads}
\label{subsecscat1} In this setting, one attaches leads to the graph
$\mathcal{G}^{(0)}$ and scattering is defined on the enlarged graph.
A lead graph $l$ is defined as a semi-infinite set of vertices $
(l,1),(l,2),\cdots $ which are connected linearly. A vertex is
identified by a double index $(l,i)$, $l$  denotes the lead, and $i$
enumerates the vertex position on the lead. The lead connectivity
(adjacency) matrix is $C^{(Lead)}_{(l,n),(l',n')}= w\delta_{l,l'}
\delta_{|n-n'|,1}\ , \ n,n'\in \mathbb{N}^+ $, where $w$ stands for
the number of parallel bonds which connect successive vertices. (All
quantities related to the leads will be denoted by the superscript
$(Lead)$). The spectrum of the lead Laplacian $\Delta^{(Lead)}_l$
and the corresponding eigenfunctions ${\bf f}=\left(f_{(l,1)
},f_{(l,2)},\cdots \right)^{\top}$ satisfy
\begin{eqnarray}
\hspace{-10mm} ( \Delta^{(Lead)}_l{\bf f})_{ (l,n) }&=& -w(f_{
(l,{n+1}) }+ f_{ (l,n-1) })+2wf_{(l,n) } =\lambda f_{(l,n)}\ \ {\rm
for}\ n>1
\ , \nonumber \\
&=&   -  w f_{(l,2)} \ \ \ \ \ \ \ \ \  +\ \  \ \ \ \ \ \ \ \ w
 f_{(l,1)} =\lambda f_{(l,1)}\ \ \ {\rm for}\ n=1 \ . \label{eq:leadlap}
\end{eqnarray}
The spectrum is continuous and supported on the spectral band
$\lambda \in [0,4w]$ (the conduction band). For any $\lambda$ in the
conduction band, there correspond two eigenfunctions which can be
written as linear combinations of counter-propagating waves:
\begin{equation}
f^{(\pm)}_{(l,n)} = \xi_{\pm}^{n-1} \ \ {\rm where} \ \ \xi_{\pm}=
1-\frac{\lambda}{2w}\pm \sqrt{\left(1-\frac{\lambda}{2w}\right
)^2-1} = {\rm e}^{\pm i\alpha(\lambda)}\ . \label{eq:free}
\end{equation}
For $\lambda >4w$, $|\xi_- |>|\xi_+| $. The reason for constructing
the leads with $w$ parallel bonds is because the conduction band can
be made arbitrarily broad. In the present application, an
appropriate choice of $w$ would be of the order of the mean valency
in the interior graph, so the spectrum of $\mathcal{G}^{(0)}$ falls
well within the conduction band.

A function which satisfies the boundary condition at $n=1$ (second
line of (\ref {eq:leadlap})) is,
\begin{equation}
f_{(l,n)} = f^{(-)}_{(l,n)} + s_l(\lambda) f^{(+)}_{(l,n)} \ ,
\end{equation}
where
\begin{equation}
s_l(\lambda) =-\frac{1-\xi_+}{1-\xi_-}\ = \xi_{+}\ \ , \ \
|s_l(\lambda)|=1 \ {\rm for}\ \lambda\in [0,4w]\ . \label{eq:sfree}
\end{equation}
The lead scattering amplitude $s_l(\lambda)$ provides the phase
gained by scattering at the end of the lead (as long as $\lambda$ is
in the conduction band). It plays here the r\^ole of the scattering
matrix which will be defined in the sequel.

Returning now to the interior graph $\mathcal{G}^{(0)}$ it is
converted into a scattering graph by attaching to its vertices
semi-infinite leads. At most one lead can be attached to a vertex,
but not all vertices should be connected to leads. Let $\mathcal{L}$
denote the set of leads, and $L=|\mathcal{L}|$. The connection of
the leads to $\mathcal{G}^{(0)}$ is given by the $V^{(0)}\times L$
 ``wiring" matrix
\begin{eqnarray}
{W}_{j,(l,1)} = \left \{
\begin{array}{l l}
1 & {\rm if}\ j\in \mathcal{V}^{(0)} \ {\rm is\  connected \ to\ }l\in \mathcal{L} \  \\
0& {\rm otherwise}
\end{array}
\right . \ . \label{eq:wiring}
\end{eqnarray}
The number of leads which emanate from the vertex $i$ is either $0$
or $1$, and is denoted by $d_i =\sum_{l\in \mathcal{L}}W_{i,(l,1)}$.
Define also the diagonal matrix  $\tilde D ={\rm diag}(d_i)$ so that
\begin{equation}
W W^\top = \tilde D \ \ \ \ ;\ \ \ \  W^\top W = I^{(L)}\ ,
 \label{eq:identities}
\end{equation}
where $I^{(L)}$ is the $L \times L$ unit matrix.

The scattering graph $\mathcal{G}$  is the union of
$\mathcal{G}^{(0)}$ and the set of leads $\mathcal{L}$. Its vertex
set is denoted by $\mathcal{V}$ and its  connectivity matrix $C$ is
given by,
\begin{eqnarray}
\forall i,j \in \mathcal{V}, \ \ : \ \ \ C_{i,j} = \left \{
\begin{array}{l l}
C^{(0)}_{i,j} & {\rm if}\ i, j\in \mathcal{V}^{(0)} \\
C^{(Lead)}_{i=(l,i),j=(l,j)} & {\rm if}\   l\in \mathcal{L}\\
w W_{i,j=(l,1)}& {\rm if}\ i\ \in \mathcal{V}^{(0)}\ {\rm and} \
l\in \mathcal{L}
\end{array}
\right . \ .
\end{eqnarray}
As is usually done in scattering theory, one  attempts to find
eigenfunctions {\bf f} of the discrete Laplacian of the scattering
graph, subject to the condition that on the leads $l=1,\cdots ,L$
the wave function consists of counter propagating waves:
\begin{equation}
f_{(l,n)} = a_l \xi_{-}^{n-1} + b_l \xi_{+}^{n-1}\ , \ n\ \ge\ 1 \ .
\label{eq:leadwave}
\end{equation}
where $a_l$ and $b_l$ are the incoming and outgoing amplitudes. They
are to be determined from the requirement that ${\bf f}$ is an
eigenfunction of the scattering graph Laplacian. It will be shown
below that this requirement suffices  to provide a linear
relationship between the incoming and outgoing amplitude. The
$L\times L$ scattering matrix $S^{(Lead)}(\lambda)$ is defined as
the mapping from the incoming to the outgoing amplitudes:
\begin{equation}
 {\bf b}=S^{(Lead)}(\lambda) {\bf a}\ .
 \label{eq:sgraph}
 \end{equation}
To compute $S^{(Lead)}(\lambda)$, consider the action of the
Laplacian on an eigenvector ${\bf f}$.
\begin{eqnarray}
\hspace {-15mm} \forall i\in \mathcal{V}^{(0)}   :   (\Delta {\bf
f})_i &=& -\sum_{j\in \mathcal{V}^{(0)}}C^{(0)}_{i,j}f_j - w
\sum_{l\in
\mathcal{L}}W_{i,(l,1)}f_{(l,1)}+\ (D_i+w d_i)f_i=\lambda f_i . \nonumber  \\
\hspace {-15mm} \forall l\in \mathcal{L}  :    (\Delta {\bf
f})_{(l,1)} &=& - w\sum_{i\in
\mathcal{V}^{(0)}}W^{\top}_{(l,1),i}f_{i}\  -w f_{(l,2)} \ \ \ \ \
+\ \ 2w f_{(l,1)}=\lambda f_{(l,1)} \ .
\nonumber\\
 (\Delta {\bf f})_{(l,n)}&=&-  w f_{(l,n+1)}\ \ \  \  -\ \ \ \
w f_{(l,n-1)}\ \  \ \ +\ \   \ 2w f_{(l,n)} =\lambda f_{(l,n)} .
 \label{eq:totlap}
\end{eqnarray}
The  equations for $ i \in \mathcal{V}^{(0)}$ (first line in (\ref
{eq:totlap}) above), can be put in a concise form:
 \begin{equation}
 \left(\Delta^{(0)} +w \tilde D-\lambda I^{(V^{(0)})}\right){\bf f}^{(V^{(0)})}=w W
 {\bf f}^{(L)}_1 \ .
 \label{eq:first}
 \end{equation}
where $I^{(V^{(0)})}$ is the unit matrix in $V^{(0)}$ dimension,
${\bf f}^{(V^{(0)})}$ is the restrictions of ${\bf f}$ to the
vertices of the interior graph $\mathcal{G}^{(0)}$ and ${\bf
f}^{(L)}_1$ is the $L$ dimensional vector with components
$f_{(l,1)},\ l=1,\cdots, L$. For $\lambda$ away from the eigenvalues
of $\Delta^{(0)} +w \tilde D$
 the $V^{(0)}\times V^{(0)}$ matrix $R^{(0)}(\lambda)$ is defined as,
\begin{equation}
R^{(0)}(\lambda) = \left(\Delta^{(0)} +w \tilde D-\lambda
I^{(V^{(0)})}\right)^{-1}\ . \label{eq:defR0}
\end{equation}
Thus,
\begin{equation}
 {\bf f}^{(V^{(0)})}=w R^{(0)} W {\bf f}^{(L)}_1\ .
 \label{eq:defR}
\end{equation}
  Substituting in the second set of equations in (\ref{eq:totlap}) and
using (\ref {eq:leadwave}),
\begin{equation}
\left (-w^2 W^{\top} R^{(0)}(\lambda) W
+(2w-\lambda)I^{(L)}\right)({\bf a}+{\bf b})=w({\bf a}\ \xi_{-}  +
{\bf b}\ \xi_{+}) .
\end{equation}
This can be easily brought into the form (\ref{eq:sgraph}). Using
the fact that $2-\lambda/w =\xi_{-}  +  \xi_{+}$ we get,
\begin{equation}
\hspace{-13mm} S^{(Lead)}(\lambda)=-\left(w W^{\top}
R^{(0)}(\lambda) W - \xi_{-}\ I^{(L)}\right)^{-1}\left(w W^{\top}
R^{(0)}(\lambda) W - \xi_{+}\ I^{(L)}\right)\ . \label{eq:sofr}
\end{equation}
This is the desired form of the scattering matrix. It has a few
important properties.

\noindent \emph{i.} As long as $\lambda$ is in the conduction band,
$\xi_-$ and $\xi_+$ are complex conjugate and unitary. Since
$W^{\top} R^{(0)}(\lambda) W$ is a symmetric real matrix,
$S^{(Lead)}(\lambda)^{\top}= S^{(Lead)}(\lambda)$ and
$S^{(Lead)}(\lambda) S^{(Lead)}(\lambda)^{\dag}=I^{(L)}$, that is,
$S^{(Lead)}(\lambda)$ is a symmetric and unitary matrix.

\noindent \emph{ii.} Once $\lambda$ is outside of the conduction
band, the $f^{(\pm)}_{(l,n)}$ are exponentially increasing or
decreasing solutions - they are the analogues of the evanescent
waves encountered in the study of wave-guides. One of the reason for
the introduction of the $w$ parallel bonds in the leads was to
broaden the conduction band and avoid the spectral domain of
evanescent waves. However, for the sake of completeness one observes
that the scattering matrix as defined above can be analytically
continued outside of the conduction band by using (\ref {eq:free})
which is valid for any $\lambda$. The $S^{(Lead)}(\lambda)$ matrix
outside the conduction band loses its physical interpretation, and
it remains symmetric but is not any more unitary. However, it is a
well defined object, and can be used in the sequel for any real or
complex $\lambda$. In the limit $\lambda \rightarrow\infty$,
$S^{(Lead)}(\lambda) \rightarrow \xi_+W^{\top} R^{(0)}(\lambda) W
\approx \left(\frac{w}{\lambda}\right)^2  I^{(L)} $.

\noindent \emph{iii.} At the edges of the conduction band
$\xi_{\pm}(\lambda =0)= 1\ ; \ \xi_{\pm}(\lambda =4w)= -1 \ $.
Substituting in (\ref{eq:sofr}) one finds that at the band edges,
$S^{(Lead)}=- I^{(L)}$.

\noindent \emph{iv.} The matrix $R^{(0)}(\lambda)$ is well defined
for $\lambda$ away from the spectrum of  $\Delta^{(0)} +w \tilde D$.
Approaching these values does not cause any problem in the
definition of $S^{(Lead)}(\lambda)$ since there $S^{(Lead)} =
-I^{(L)}$. However, for sufficiently large  $w$ the singularities of
$R^{(0)}(\lambda)$ can be separated away from the domain where the
spectrum of $\mathcal{G}^{(0)}$ is supported.

\noindent \emph{v.} The resonances are defined as the poles of the
scattering matrix in the complex $\lambda$ plane. They are the
solution of the equation
\begin{equation}
z_{res}(\lambda)=\det\left(wW^{\top} R^{(0)}(\lambda) W - \xi_{-}\
I^{(L)}\right)=0\ .
\end{equation}
The point $\lambda=0$ is not a pole since $S^{(Lead)}(\lambda=0)= -
I^{(L)}$.

\noindent \emph{vi.} Finally, it might be instructive to note that
the matrix $R^{(0)}(\lambda)$ is closely related to the discrete
analogue of the Dirichlet to Neumann map. This can be deduced from
the following construction: add to each vertex $i\in
\mathcal{V}^{(0)}$ a new auxiliary vertex $\tilde i$ connected
exclusively to $i$. (Here we use $w=1$ to make the analogy clearer).
Write the discrete Laplacian for the new graph, and solve
$(\Delta-\lambda I){\bf \tilde f}=0$, where ${\bf \tilde f}$ is a
$2V^{(0)}$ dimensional vector, the first $V$ entries correspond to
the original vertices, and the last $V$ entries correspond to the
auxiliary vertices : ${\bf \tilde f}=({\bf f},{\bf g})^{\top}$.
Assuming that the values $g_{\tilde i}$ on the auxiliary vertices
are given, the entries in ${\bf f}$ can be expressed as $ {\bf f}=
R(\lambda) {\bf g}$, where $R(\lambda)$ as defined in
(\ref{eq:defR0}). To emphasize the connection to the Dirichlet to
Neumann map, define ${\bf \psi} = \frac{1}{2}({\bf g}+{\bf f})$ (the
``boundary function") and $\partial{\bf \psi} = ({\bf g}-{\bf f})$
(the ``normal derivative") then,
\begin{equation}
\partial{\bf \psi}= M(\lambda) {\bf \psi} \ \ \ ; \ \ \
M(\lambda)=2(I^{(V^{(0)})}+R(\lambda))^{-1}(I^{(V^{(0)})}-R(\lambda))\
.
\end{equation}
The Dirichlet to Neumann map is defined also in other applications
of graph theory, see e.g., \cite{CU91}.

\subsection{Scattering into dangling bonds (evolution operator approach)}
\label{subsecscat2} An alternative construction of a scattering
matrix follows naturally from the study of the evolution operator
$U^{(0)}(\lambda)$ of the interior graph.  It is a unitary matrix of
dimension $2E^{(0)}$ which will be reviewed in the following
paragraphs for the sake of completeness. (Details can be found in
\cite{US07,Mizuno08}. See also \cite{Novikov,
Cataneo,Exner,Kuchment}, and references cited therein).

Let ${\bf f}=(f_1,\cdots,f_{V^{(0)}})$ denote an eigenvector of the
graph Laplacian  $\Delta^{(0)}$ (\ref{eq:lap0}), corresponding to an
eigenvalue $\lambda$. The bond $b$ connecting the vertices $i$ and
$j$, will be denoted by $b =(i,j)$.  To each bond $b$ one associates
a \emph{bond wave function}
\begin{equation}
\psi_b(x) = a_b\ {\rm e}^{i\frac{\pi}{4}x} + \hat a_{ b}\ {\rm
e}^{-i\frac{\pi}{4}x} \ \ \ ,
 \ \ x\in \{\pm1\}
 \label{eq:wfunction}
 \end{equation}
subject to the condition
\begin{equation}
\psi_b(1)= f_i \ \ \ , \ \ \ \psi_b(-1)= f_j  \ . \label{eq:bcond}
 \end{equation}
Consider any vertex indexed by $i$ of degree (valency) $v_i$, and
the bonds $(b_1,b_2, ...b_{v_i})$ which emanate  from $i$. The
corresponding bond wave functions have to satisfy three requirements
in order to form a proper eigenvector of $\Delta^{(0)}$.

\noindent {\it I. Uniqueness}: The value of the eigenvector at the
vertex $i$, $f_i$, computed in terms of the bond wave functions is
the same for all the bonds emanating from $i$. The following $v_i-1$
independent equalities express this requirement.
 \begin{equation}
 \hspace{-15mm}
 a_{b_1}\ {\rm e}^{i\frac{\pi}{4}} + \hat a_{ b_1}\ {\rm
e}^{-i\frac{\pi}{4}} = a_{b_2}\ {\rm e}^{i\frac{\pi}{4}} + \hat a_{
b_2}\ {\rm e}^{-i\frac{\pi}{4}} =\ \cdots \  = a_{b_{v_i}}\ {\rm
e}^{i\frac{\pi}{4}} + \hat a_{ b_{v_i}}\ {\rm e}^{-i\frac{\pi}{4}}\
.
 \label {eq:uniq}
\end{equation}

\noindent {\it II. ${\bf f}$ is an eigenvector of $\Delta^{(0)}$}  :
At the vertex $i$, $\sum_{j=1}^{v_i} {\Delta^{(0)}}_{i,j}f_j=\lambda
f_i$. In terms of the bond wave functions this reads,
\begin{equation}
\hspace{-10mm}
 -\sum_{l=1}^{v_i}\left [a_{b_l}\ {\rm
e}^{-i\frac{\pi}{4} } + \hat a_{ b_l}\ {\rm e}^{+i\frac{\pi}{4}
}\right ]
 =(\lambda - v_i)\ \frac{1}{v_i} \sum_{m=1}^{v_i}
\left [ a_{b_m}\ {\rm e}^{i\frac{\pi}{4}} + \hat a_{b_m}\ {\rm
e}^{-i\frac{\pi}{4}}\right ]\ .
 \label{eq:lapleq}
 \end{equation}
 Together, (\ref{eq:uniq}) and (\ref{eq:lapleq}) provide
$v_i$ homogeneous linear relations  between  the $2v_i$ coefficients
$a_{b_m},\hat a_{b_m}$. It is convenient to introduce at this point
the following notation: Let $d=(j,i),\ i,j\in \mathcal{V}^{(0)}$
denote a directed bond pointing from $i$ to $j$. Then $o(d)$
($t(d)$) stand for its origin $i$ (terminus $j$) vertices,
respectively. From now on, the amplitudes $a_{b_m}$ and $\hat
a_{b_m}$ which refer to propagating on the bond $b_m$ in opposite
directions will be denoted by $a_d$ and $a_{\hat d}$ where $\hat d$
is the directed bond inverse to $d$.  Using (\ref{eq:uniq}) and
(\ref{eq:lapleq}), the outgoing coefficients are expressed in terms
of the incoming ones,
 \begin{equation}
 \label{eq:scatmat}
a_d =\sum_{d'\ :\ t(d')=i}  \sigma^{(i)}_{d,d'}(\lambda) \ a_{d'}\ \
\
   \ \forall \   d \ :\ o(d)=i\ ,
\end{equation}
where,
\begin{equation}
 \label{eq:scatmatdetail}
\sigma^{(i)}_{d,d'}(\lambda )= i\left(\delta_{\hat d,
d'}-\frac{2}{v_i}\frac{1}{1-i(1-\frac{\lambda}{v_i})} \right) \ .
\end{equation}
The \emph{vertex scattering matrices} $\sigma^{(i)}(\lambda)$ are
the main building blocks of the present approach.  A straight-
forward computation shows that for real $\lambda$ the vertex
scattering matrices are unitary and symmetric matrices.

\noindent {\it III. Consistency} : The linear relation between the
incoming and the outgoing coefficients (\ref {eq:scatmat}) must be
satisfied simultaneously at all the vertices. However, a directed
bond $(i,j)$ when observed from the vertex $j$ is \emph{outgoing},
while when observed from $i$ it is \emph{incoming}. This consistency
requirement is implemented  by introducing the \emph{Evolution
Operator} $U_{d'.d}(\lambda)$ in the $2E^{(0)}$ dimensional space of
directed bonds,
\begin{equation}
 U_{d',d}(\lambda) = \delta_{t(d),o(d')}\
 \sigma^{(t(d))}_{d',d}(\lambda)\ .
 \label{eq:umatrix}
\end{equation}
The evolution operator is unitary $ U\ U^{\dagger} = I^{(2E^{(0)})}$
for $\lambda\in \mathbb{R}$ due to the unitarity of its constituents
$\sigma ^{(i)}$. Denoting by $\bf a$ the $2E^{(0)}$ dimensional
vector of the directed bonds coefficients $a_d$ defined above, the
consistency requirement reduces to,
\begin{equation}
U(\lambda)\ {\bf a} = {\bf a}\ .
 \label{eq:consisu}
\end{equation}
This can be satisfied only for those values of $\lambda$ for which
\begin{equation}
\det\left (I^{(2E^{(0)})}-U(\lambda)\right )\ = \ 0 \ .
 \label{eq:secu}
\end{equation}
This result can be interpreted in the following way.  The evolution
operator $U(\lambda)$ is defined for any $\lambda$, and it maps the
$2E^{(0)}$ dimensional vector space of amplitudes ${\bf a}$ to
itself. The spectrum of $\Delta^{(0)}$ is identified as those values
of $\lambda$ for which there exist vectors which are {\it
stationary} under the action of the mapping.

To construct a scattering operator for which $\mathcal{G}^{(0)}$ is
the ``interior", add to the vertex set $\mathcal{V}^{(0)}$ another
set of vertices denoted by $\mathcal{L}$ with $L=|\mathcal{L}|$.
Connect $\mathcal{L}$ to a subset of $\mathcal{V}^{(0)}$ such that
each $l\in \mathcal{L}$ is connected to a single vertex in
$\mathcal{V}^{(0)}$, while a vertex in $\mathcal{V}^{(0)}$ can be
connected to several vertices in $\mathcal{L}$. The vertices in
$\mathcal{V}^{(0)}$ connected to $\mathcal{L}$ will be referred to
as boundary vertices. The $V^{(0)}\times L$ ``wiring" matrix is
defined similarly to the previous definitions (see
(\ref{eq:wiring})),
\begin{eqnarray}
{W}_{j,l} = \left \{
\begin{array}{l l}
1 & {\rm if}\ j\in \mathcal{V}^{(0)} \ {\rm is\  connected \ to\ }l\in \mathcal{L} \  \\
0& {\rm otherwise}
\end{array}
\right . \ . \label{eq:wiringu}
\end{eqnarray}
The number of ``dangling bonds" which emanate from the vertex $i$ is
$d_i=\sum_{l\in \mathcal{L}}W_{i,l}\ \ $ and  $d_i$ can take any
integer value or $0$. The diagonal matrix with elements $d_i$ will
be denoted by $\tilde D$ and the identities (\ref{eq:identities})
hold. The new graph $\tilde \mathcal{G}$ is of  cardinality $\tilde
V =V^{(0)}+L$ and it consists of the interior graph with $L$
dangling bonds attached.

The evolution operator for  $\tilde \mathcal{G}$ can be written in
block form as
\begin{eqnarray}
\tilde U = \left (
\begin{array}{l l l}
 \Sigma &\Omega & 0  \\
0 & 0 & \rho ^{(out)}\\
\Omega^{tr} & \rho ^{(in)} &0
\end{array}
\right ) \ , \label{eq:Utilde}
\end{eqnarray}
which is a $2(E^{(0)}+L)\times2(E^{(0)}+L)$ unitary matrix  arranged
in the following way. The first $2E^{(0)}$ rows and columns are
labeled by the indices of the directed bonds which belong to the
interior graph $\mathcal{G}^{(0)}$. The last $2L$ rows and columns
are labeled by the indices of the directed dangling bond. The first
$L$ correspond to \emph{incoming} bonds pointing from $\mathcal{L}$
to $\mathcal{V}^{(0)}$, the other $L$ indices correspond to
\emph{outgoing} bonds from $\mathcal{V}^{(0)}$ to $\mathcal{L}$. The
$2E^{(0)}\times 2E^{(0)}$ upper left block, denoted by $\Sigma$ is
obtained by modifying the evolution operator for the interior graph:
the vertex scattering matrices (\ref{eq:scatmatdetail}) are modified
by replacing the original valency $v_i$ by the modified valencies
${\tilde v}_i = v_i +d_i$. The rectangular $ 2E^{(0)}\times L$
matrix $\Omega$ consists of elements of the vertex scattering
matrices between incoming dangling bonds and directed bonds in
$\mathcal{G}^{(0)}$ which emanate from boundary vertices in
$\mathcal{V}^{(0)}$. $\Omega ^{tr}$ provides the scattering matrix
elements for the time reversed transitions, from the directed bonds
in $\mathcal{G}^{(0)}$ to the outgoing dangling bonds. The $L\times
L$ matrix $\rho^{(out)}$ consists of matrix elements from outgoing
to incoming dangling bonds which occur at the dangling vertices
$\mathcal{L}$. Using (\ref{eq:scatmatdetail}) with $v_l=1$, we get
\begin{equation}
\rho^{(out)}_{d',d}
 =-i\frac{1+i(1-\lambda)}{1-i(1-\lambda)}
 \delta_{o(d'),t(d)}W_{o(d'),t(d)}
 \label{eq:rhoout}
 \end{equation}
The $L\times L$ matrix $\rho^{(in)}$  consists of matrix elements
from incoming to outgoing dangling bonds which occur at the boundary
vertices. They are computed using (\ref{eq:scatmatdetail}) with
${\tilde v}_i = v_i +d_i$. Note: the number of dangling bonds
connected to the same vertex can exceed 1, and in these cases
scattering to other dangling bonds may occur, as long as both the
incoming and outgoing bonds are connected to the same vertex of
$\mathcal{G}^{(0)}$ . The four zero blocks correspond to transitions
between bonds which do not follow each other. In particular, the
second row of blocks in (\ref{eq:Utilde}) has two zero entries,
expressing the fact that ingoing dangling bonds cannot follow
neither interior bonds nor  any ingoing dangling bonds.

The unitarity of $\tilde U$ ensures the following relations between
its components:
\begin{eqnarray}
\label{eq:unitutil}
 a. \ \ \ &\ & \Sigma \Sigma ^{\dag} +\Omega \Omega{^\dag} =
I^{(2E^{(0)})}\nonumber \\
 b. \ \ \ &\ &\rho^{(out)}{\rho ^{(out)}}^{\dag}=I^{(L)}\\
 c. \ \ \ &\ &\Omega^{tr}{\Omega^{tr}}^{\dag} +
\rho^{(in)}{\rho^{(in)}}^{\dag} = I^{(L)}\nonumber \\
 d. \ \ \ &\ &\Sigma{\Omega^{tr}}^{\dag} +\Omega {\rho^{in}}^{\dag} =
 0\nonumber
\end{eqnarray}

The evolution operator $\tilde U$ acts in the $2E^{(0)}+2L$
dimensional space of amplitude vectors. Denote the $2E^{(0)}$
amplitudes which refer to directed bonds in $\mathcal{G}^{(0)}$ by
${\bf a}$ and the amplitudes associated with the incoming/outgoing
dangling bonds by ${\bf b}^{(-/+)}$ respectively. The consistency
condition (\ref{eq:consisu}) which determines the spectrum of
$\tilde \mathcal{G}$ reads now,
\begin{eqnarray}
\left (
\begin{array}{l l l}
\Sigma &\Omega & 0  \\
0 & 0 & \rho ^{(out)}\\
\Omega^{tr} & \rho ^{(in)} &0 \end{array} \right ) \left (
\begin{array}{l }
 {\bf a}  \\
 {\bf b^{(-)}}\\
{\bf b^{(+)}}
\end{array}
\right )=\left (
\begin{array}{l }
 {\bf a}  \\
 {\bf b^{(-)}}\\
{\bf b^{(+)}}
\end{array}
\right )
 \ . \label{eq:Utildeev}
\end{eqnarray}
To define a scattering matrix, one has to obliterate the requirement
that scattering occurs at the dangling vertices. This is achieved by
replacing $\rho^{(out)}$ by a $0$ block in $\tilde U$. By doing so,
the rank of $\tilde U$ is reduced by $L$. Thus, the resulting
equations do not determine a spectrum. Rather, they can be solved
for every value of $\lambda$, and yield a linear relation between
the amplitudes of the outgoing and incoming dangling bonds,
\begin{eqnarray}
{\bf b}^{(+)} &=&\left ( \rho^{(in)}+ \Omega^{tr}\
(I^{(2E^{(0)})}-\Sigma)^{-1}\Omega \right ) {\bf b}^{(-)} \ \ \
\doteq\
\ S^{(D)} (\lambda){\bf b}^{(-)}\ \nonumber \\
 S^{(D)} (\lambda)&=&  \rho^{(in)}(\lambda)+ \Omega^{tr}(\lambda)\
(I^{(2E^{(0)})}-\Sigma(\lambda))^{-1}\Omega(\lambda)  \ .
 \label{eq:sofu}
\end{eqnarray}
The $L\times L$ matrix $S^{(D)}(\lambda)$ (The superscript ``D"
stands for Dangling) provides the scattering amplitudes  between
incoming and outgoing dangling bonds. It is  meromorphic in
$\lambda$, has a finite number of poles in the lower half plane, and
is unitary on the real $\lambda$ axis. These statements can be
easily checked by using the explicit form of the matrix elements of
$\tilde U(\lambda)$ and the identities (\ref{eq:unitutil}). The main
advantage of $S^{(D)}(\lambda)$ over the matrix
$S^{(Lead)}(\lambda)$ defined in (\ref{eq:sofr}) is that it depends
solely on properties of the interior graph, and its definition is
free from restrictions or special features due to the properties of
the leads.

\section{The exterior-interior duality}
\label{sec:ext-int}
 In the present section, two versions of the
exterior interior duality  will be formulated corresponding to the
two ways by which scattering was defined above. The conversion of a
graph into a scattering graph can be done in many ways - the
vertices connected to leads or to  dangling bonds can be chosen
arbitrarily. It will be shown here, that any scattering matrix can
be used to extract information about the spectrum of the ``interior"
graph. However, there exists a unique construction which provides
the entire spectrum of the interior. Namely, when all the vertices
are connected to leads (in the first setting) or dangling bonds (one
per vertex,  in the second setting): $L=V^{(0)}$ and
$W_{i,l}=I^{(V^{(0)})}$. Under these conditions the secular
equations
\begin{equation}
\det (I^{(V^{(0)})}- S(\lambda)) =0  ,\ \  {\rm with} \ \ S =
\xi_{-} S^{(Lead)}\ {\rm or}\ S= S^{(D)}
\end{equation}
provide the complete spectrum of the interior graph with
multiplicities.

\subsection{The exterior-interior duality for scattering to leads}
 \label{subsec:ext-int-leads}
Equations (\ref {eq:totlap}) form the basis of this discussion. They
can be solved for any value of $\lambda$ and they provide the
building blocks for constructing the scattering matrix
$S^{(Lead)}(\lambda)$. Consider the subset (\ref{eq:first}) of these
equations. If for a given $\lambda$  the corresponding vectors ${\bf
f}^{(V^{(0)})}$ and ${\bf f}_1^{(L)}$ which solve (\ref {eq:totlap})
also satisfy
 \begin{equation}
 \tilde D{\bf f}^{(V^{(0)})}= W {\bf f}_1^{(L)}\ \ {\rm and} \ \ {\bf f}^{(V^{(0)})}\ \ne \ 0
 \end{equation}
then, ${\bf f}^{(V^{(0)})}$ is an eigenvector of the interior graph
Laplacian with an eigenvalue $\lambda$. If the leads are not
connected to all the vertices, that is $L<V^{(0)}$, the  kernel of
$\tilde D$ is not empty. Hence, it is possible that the condition
$\tilde D{\bf f}^{(V^{(0)})} = W {\bf f}_1^{(L)} $ is satisfied
trivially by ${\bf f}^{(V^{(0)})}$ which is in the right kernel of
$\tilde D$, and ${\bf f}_1^{(L)}=0$. Such cases have to be excluded
from the following derivation and therefore, in general, the
equation to be derived below provides only a sufficient spectral
condition. However, if one connects every vertex to a lead, this
problem does not arise, and a proper secular equation can be
derived, as will be shown in the sequel.

As long as $\lambda$ is away from the singularities of
$R^{(0)}(\lambda)$, one can proceed and obtain from $\tilde D{\bf
f}^{(V^{(0)})}=  W {\bf f}_1^{(L)} $
\begin{equation}
(w\tilde D R^{(0)} -I^{(V^{(0)})})W {\bf f}_1^{(L)}=0 \ .
\label{eq:conditionlead}
\end{equation}
Using (\ref{eq:identities}) one finds that (\ref{eq:conditionlead})
is equivalent to requiring a non trivial solution for the equation
\begin{equation}
W \left (w W^{\top} R^{(0)}(\lambda) W- I^{(L)}\right ){\bf
f}_1^{(L)}=0\ .
\end{equation}
 Thus, it is sufficient but not necessary
for $\lambda$ to be in the spectrum if it is a zero of the function
\begin{equation}
z_R(\lambda) = \det \left (w W^{\top} R^{(0)}(\lambda) W-
I^{(L)}\right ) \ . \label{eq:zetl}
\end{equation}
However, when all the vertices are connected to leads,
$W=I^{(L)}=I^{(V^{(0)})}$ , $\tilde D = I^{(V^{(0)})}$ and
$R^{(0)}(\lambda) = \left(\Delta^{(0)} -(\lambda-w)
I^{(V^{(0)})}\right)^{-1}$. Substituting in (\ref{eq:zetl}), one
gets
\begin{equation}
z_R(\lambda) = (-1)^{V^{(0)}}\frac{\det \left ( \Delta^{(0)}-\lambda
I^{(V^{(0)})}\right )}{\det \left ( \Delta^{(0)}-(\lambda-w)
I^{(V^{(0)})}\right )} \ . \label{eq:zetlcom}
\end{equation}
At this point the importance of the parameter $w$ becomes clear: The
spectrum of the Laplacian coincides with the zeros of $z_R(\lambda)$
only if there are no accidental overlaps with the poles of
$R^{(0)}(\lambda)$. Thus, one should choose $w$ so that the spectrum
of $(\Delta^{(0)} +w I^{(V^{(0)})})$ exceeds the maximum of the
spectrum of $\Delta^{(0)}$. For $v$ regular graphs taking  $w>2v$ is
sufficient. Using (\ref{eq:sofr}) one writes
\begin{equation}
 wW^{\top} R^{(0)}(\lambda) W= \left(
 I^{(L)}+S^{(Lead)}(\lambda)\right )^{-1}(\xi_+\ I^{(L)}+\xi_-\ S^{(Lead)}(\lambda)).
\end{equation}
Substituting in (\ref{eq:zetl}) yields the secular equation
\begin{equation}
\hspace{-10mm} z_{Ld}(\lambda)=\det \left (I^{(L)} -
s_l^{-1}(\lambda) S^{(Lead)}(\lambda)\right )\ = \det \left (I^{(L)}
- \xi_- S^{(Lead)}(\lambda)\right )\ = \ 0 \ . \label{eq:zets}
\end{equation}
This is the desired result, since it determines the spectrum of the
interior graph in terms of the scattering matrix
$S^{(Lead)}(\lambda)$ and the free lead scattering matrix
$s_l(\lambda)$ (\ref{eq:sfree}). It is analogous both in form and in
content to the exterior-interior secular equation which was
introduced in the introduction (\ref {eq:originalEID}).

\subsection {Exterior -Interior duality for scattering to dangling bonds}
\label{subsec:e-i dangling}

The scattering matrix $S^{(D)}(\lambda)$ will be used to compute the
spectrum of the interior graph $\mathcal{G}^{(0)}$. For this
purpose, consider the action of the Laplacian of the extended graph
$\tilde \mathcal{G}$. Denote by $f_j$ the components of an
eigenvector on a vertex $j\in \mathcal{V}^{(0)}$, and by $g_k$ the
component of the same eigenvector on a dangling vertex $k\in
\mathcal{L}$.  Then,
\begin{equation}
\forall\ i \ :\ \  -\sum_{j\in \mathcal{G}^{(0)}}C_{i,j}f_j
-\sum_{l=1}^{L} W_{i,l}g_l +(v_i+d_i)f_i=\lambda f_i \ .
\end{equation}
$\lambda$ is an eigenvalue of the interior graph, corresponding to
an eigenvector ${\bf f}=(f_1,\cdots,f_{V^{(0)}})$, if for every
$i\in \mathcal{V}^{(0)} $
\begin{equation}
\sum_{l=1}^{L}W_{i,l} g_l =d_i f_i \ , \label{eq:conditiondang}
\end{equation}
and ${\bf f}\ne 0$. Consider a single  boundary vertex $j$ with
$d_j\ne 0$. Expressing the eigenfunction on the dangling bonds in
terms of the incoming and outgoing amplitudes (\ref {eq:wfunction}):
\begin{eqnarray}
\forall k\ : \ W_{j,k}=1\ , \ \ &f&_j = b^{(-)}_k \ {\rm
e}^{-i\frac{\pi}{4}} + b^{(+)}_k\ {\rm
e}^{+i\frac{\pi}{4}} \ ,\nonumber\\
 &g&_k = b^{(-)}_k \ {\rm
e}^{+i\frac{\pi}{4}} + b^{(+)}_k\ {\rm e}^{-i\frac{\pi}{4}} \ .
 \label{eq:wfuncdgbond}
 \end{eqnarray}
The upper line above imposes $d_j-1$ linear relations between the
amplitudes. Substituting  (\ref{eq:wfuncdgbond}) in (\ref
{eq:conditiondang}) results in one more linear relation. Thus, one
gets for each $j$ exactly $d_j$ linear relations between the
incoming and outgoing amplitudes on the dangling bonds connected to
the interior vertex $j$. Solving them for each $j$ separately and
combining the individual relationships gives:
 \begin{equation}
\textbf{b}^{(+)}  = S^{(0)}\textbf{b}^{(-)} \ \ \ ;\ \ \
S^{(0)}_{k,l}= \sum_{j\in \mathcal{V}^{(0)}}W_{j,k}W_{j,l} \left
(\frac{1-i}{d_j}+i\delta_{k,l}\right )\ .
 \label{eq:sefes}
\end{equation}
This  $\lambda$ independent matrix of dimension $L\times L$ is bloc
diagonal, with blocs of dimensions $d_j$. It is easy to check that
each block (and hence the entire matrix) is unitary and symmetric.
When only one dangling bond is connected to a vertex, $d_j=1$ and
the corresponding  $S^{(0)}_{k,k}=1$.

The requirement (\ref {eq:sefes}) has to be combined now with the
relation (\ref {eq:sofu}) which is valid for all $\lambda$. This
implies that $\lambda$ is an eigenvalue of the Laplacian on the
interior graph if there exists a non trivial solution for the
equation
\begin{equation}
{\bf b}^{(+)} =S^{(D)}(\lambda)\left(S^{(0)}\right)^{\dag}{\bf
b}^{(+)}\ ,
\end{equation}
in other words, if $\lambda$ is a zero of the secular function:
\begin{equation}
z_D(\lambda) = \det \left (I^{(L)}-S^{(D)}(\lambda)\  \left
(S^{(0)}\right ) ^{\dag}\right )\ .
 \label{eq:seculss}
\end{equation}
The derivation above follows from the requirement that (\ref
{eq:conditiondang}) is satisfied for ${\bf f}\ne 0$. This can be
guaranteed only when a single dangling bond is connected to each
vertex. In this case, $S^{(0)}=I^{(V_0)}$  and the  secular equation
(\ref {eq:seculss}) expresses the exterior - interior duality in the
present setting.

\section {Examples and Applications}\label {sect:applications}

In this section a  few examples and applications which will
illustrate the exterior - interior duality will be discussed.

\subsection {Scattering on a single lead connected to a graph}
\label{subsect:single lead}

Consider a single lead connected to one of the vertices of
$\mathcal{G}^{(0)}$ which is denoted by $i=1$. Following (\ref
{eq:defR0}, \ref{eq:sofr}),
\begin{eqnarray}
R_{1,1}(\lambda) &\doteq & W^{\top} R(\lambda) W =
\frac {G_{1,1}(\lambda)}{1+w G_{1,1}(\lambda)}\ , \nonumber \\
G_{1,1}(\lambda)&\doteq & \left ( (\Delta^{(0)} - \lambda
I^{(0)})^{-1}\right)_{1,1}\
=\sum_{r=1}^{V^{(0)}}\frac{|f^{(r)}_1|^2}{\lambda_r-\lambda} .
\end{eqnarray}
Clearly $G(\lambda)$ is the Green function (resolvent) of the graph
Laplacian. The one dimensional scattering matrix reads,
\begin{equation}
S^{(Lead)}(\lambda) = -\frac{wR_{1,1}-\xi_+}{wR_{1,1}-\xi_-}\ ,
\end{equation}
and the secular equation:
\begin{equation}
z_{Ld}(\lambda) =-\frac{1+\xi_-}{wR_{1,1}(\lambda)-\xi_-}\
 \frac{1}{1+wG_{1,1}(\lambda)} \label{eq:zetlsingle}
\end{equation}
This function has no poles for $\lambda$ in the conduction band, and
it vanishes at the spectrum of $\Delta^{(0)}$ (= the poles of the
Green function), provided that the corresponding residues do not
vanish. This cannot be guaranteed since the component of an
eigenvector can vanish on any number of vertices. Such eventualities
cannot be excluded and therefore the proper secular equation has to
be constructed by connecting all the vertices to leads.

\subsection{Composition}\label{susec:composition} Consider two
graphs $\mathcal{G}^{(1)}$ and $\mathcal{G}^{(2)}$ with vertex sets
$\mathcal{V}^{(i)}$ of cardinality $V^{(i)}$ and connectivity
matrices $C^{(i)}$ , $i=1,2$. Connect the two graphs by an arbitrary
number of leads $(j,i), \ i\in \mathcal{V}^{(1)} , \ j\in
\mathcal{V}^{(2)}$ but avoid parallel edges. Denote the number of
connecting edges by $L$. The connection between the two graphs is
given by the matrix:
\begin{eqnarray}
{\tilde C}_{i,j} = \left \{
\begin{array}{l l}
   1 & {\rm if}\ i\in \mathcal{V}^{(1)} , \ j\in
\mathcal{V}^{(2)}\ {\rm are\ connected.} \\
0 & {\rm otherwise.}
\end{array}
\right .
\end{eqnarray}
The purpose is to obtain a secular equation for the composite graph
$G^{(0)}=\mathcal{G}^{(1)} \cup \mathcal{G}^{(2)}$ with connectivity
matrix
\begin{eqnarray}
{\tilde C}^{(0)} = \left  (
\begin{array}{l l}
   C^{(1)} & {\tilde C}  \\
{\tilde C}^{\top} & C^{(2)}
\end{array}
\right )\ .
\end{eqnarray}
Consider the graph $\mathcal{G}^{(1)}$ say, and regard for the
moments the bonds which connect it to $\mathcal{G}^{(2)}$ as
dangling bonds. The corresponding wiring matrix is $\tilde C$, and
one can write an $L$ dimensional scattering matrix $S^{(1)}(\lambda
)$ for this system, by using the ``dangling bond" construction.
Similarly, the scattering matrix $S^{(2)}(\lambda)$ which is also
$L$ dimensional can be written for $\mathcal{G}^{(2)}$. Denote by
${\bf a}$ and ${\bf b}$ the $L$ dimensional vectors of amplitudes
for incoming waves to $\mathcal{V}^{(1)}$. The consistency
requirement of section (\ref{subsecscat2}) requires that
\begin{equation}
{\bf b} = S^{(1)}(\lambda) {\bf a}\ \ ;\ \  {\bf a} =
S^{(2)}(\lambda) {\bf b}\ \ \Rightarrow\ \ {\bf b} =
S^{(1)}(\lambda)S^{(2)}(\lambda) {\bf b}\ ,
\end{equation}
so that a sufficient condition for $\lambda$ to be in the spectrum
of $\mathcal{G}^{(0)}$ is that it is a solution of the equation
\begin{equation}
\det(I^{(L)}-S^{(1)}(\lambda)S^{(2)}(\lambda))= 0\ .
\end{equation}
This  is the analogue of the example given in the introduction of
the exterior - interior duality in $\mathbb{R}^2$.

As a simple corollary of this result, one can study the effect of
adding a single vertex $m$ of valency $v_{m}$ to a given graph of
cardinality $V$. Assuming that the new vertex is connected to the
vertices $1,\cdots, v_{m}$, one computes the corresponding
$v_{m}\times v_{m}$ scattering matrix $S^{(V)}$ from the original
graph. $S^{(V)}$ plays the r\^ole of $S^{(1)}(\lambda)$ above.  For
$S^{(2)}(\lambda)$ one uses the vertex scattering matrix
$\sigma^{(m)}$ (\ref {eq:scatmatdetail}). The resulting secular
equation reads now
\begin{equation}
\det(I^{(v_{V+1})}-S^{(V)}(\lambda)\sigma^{(m)}(\lambda))  =0\ .
\end{equation}

The same problem can be addressed in a different way. With the same
construction in mind, one can compute the $R(\lambda)$ matrix (with
$w=1$). Connecting together all the $v_m$ dangling bonds to a single
vertex, and setting the value of the wave function to $f_m$ at the
vertex $m$, one finds for the values of the function on the m
connected vertices:
\begin{equation}
f_j = f_m \sum_{k=1}^{v_m}R(\lambda)_{j,k}  .
\end{equation}
At the same time, to be an eigenvalue ${\bf f}$ must satisfy
\begin{equation}
- \sum_{j=1}^{v_m} f_j + v_m f_m =\lambda f_m .
\end{equation}
Thus, the secular equation for the spectrum of the enlarged graph
reads
\begin{equation}
\lambda  = v_m - \sum_{k,j=1}^{v_m}R(\lambda)_{j,k}  .
\end{equation}
It is hoped that the few simple illustrations given above illustrate
the advantages and potential applications of the exterior-interior
duality in the present context.

 \section*{Acknowledgments}

\noindent  I would like to thank Rami Band and Amit Godel for
critical comments and suggestions, Dr Iwao Sato for carefully
reading the manuscript, to Yehonatan Elon and Idan Oren for helpful
discussions and to Professor Marco Marletta for his illuminating
comments on the Dirichlet to Neumann map. This work was supported by
the Minerva Center for non-linear Physics, the Einstein (Minerva)
Center at the Weizmann Institute and the Wales Institute of
Mathematical and Computational Sciences) (WIMCS). Grants from GIF
(grant I-808-228.14/2003), and BSF (grant 2006065) are acknowledged.

\end{document}